\documentstyle[amssymb,aps,twocolumn,epsf,epsfig]{revtex}

\begin{document}
\title{Fast computation of the Kohn-Sham susceptibility of large systems}
\author{D. Foerster}
\address{CPMOH, Universit\'{e} Bordeaux 1\\
351, cours de la Lib\'{e}ration\\
33405 TALENCE Cedex, France}
\date{\today}
\maketitle

\begin{abstract}
For hybrid systems, such as molecules grafted onto solid surfaces, the
calculation of linear response in time dependent density functional theory
is slowed down by the need to calculate, in $\sim $ $N^{4}$ operations, the
susceptibility of $N$ non interacting Kohn-Sham reference electrons. We
observe that this susceptibility can be calculated $N$ times faster, with a
precision of $\sim \left( \Delta \omega /\varepsilon \right) ^{2}$, where $%
\Delta \omega $ is the width of the frequency intervals and $\varepsilon $ a
displacement off the real frequency axis. By itself or in combination with
previously known procedures of accelerating such calculations, our procedure
should significantly facilitate the calculation of TDDFT response and
optical spectra of hybrid systems.
\end{abstract}

\vspace{18pt}%
%
{}{\it Motivation}. Time dependent density functional theory (TDDTFT) has
proved to be very successful in determining optical properties of molecules
and clusters, for a review see \cite{TDDFTSuccess}. Therefore it would be
interesting, to study within TDFFT linear response, the simplest version of
this theory, also hybrid materials, such as dyes bound to semiconductor or
metal surfaces. Such materials exhibit interesting phenomena, for example
charge transfer upon photo excitation in the Gr\"{a}tzel cell \cite{Graetzel}
or increased fluorescence for organic molecules chemisorbed onto noble metal
clusters \cite{Kamat}.

Density functional algorithms based on localized orbitals, such as the
siesta package \cite{siesta} have been used to study the ground state of
systems as large as DNA chains \cite{DNA}, but the computation of optical
spectra and properties of excited states of hybrid systems in such schemes
appears still to be very difficult.

The present note is motivated by recent progress in the calculation of
optical spectra in TDDFT linear response \cite{BlaseOrdejon} using localized
orbitals and the observation that even this recent advance is not sufficient
to deal with hybrid systems. To model such systems one must deal with a slab
of a few layers of solid onto which a molecule may be adsorbed, without the
benefit of any symmetries, and with $N$, the number of orbitals, at least of
the order of a few thousand. In the local orbital approach of \cite
{BlaseOrdejon} the needed CPU time grows rather steeply, as $\sim $ $N^{4}$,
with $N$ and paradoxically, most of the CPU time is actually spent on the
charge susceptibility of the non interacting Kohn-Sham reference fermions.

Here we focus on the Kohn-Sham charge susceptibility and describe a simple
way to calculate it $N$ times faster than by straightforward computation, in 
$\sim $ $N^{3}$ rather than $\sim $ $N^{4}$ operations, with a relative
precision of $\sim \left( \Delta \omega /\varepsilon \right) ^{2}$, where $%
\Delta \omega $ is the spacing of frequencies, and we also study the
performance of this technique by applying it to a random hamiltonian.

The equations of \ TDDFT linear response {\cite{Gross}} are based on the
definition of the time dependent Kohn-Sham potential $V_{KS}(r,t)$ of non
interacting reference electrons: 
\begin{equation}
V_{KS}(r,t)=V_{ext}(r,t)+\int dr^{\prime }\frac{n(r^{\prime },t)}{%
|r-r^{\prime }|}+V_{xc}(n(r,t))\text{ }
\end{equation}
where $V_{ext}$ is the external potential and $V_{xc}$ the exchange
correlation potential which reduces to an ordinary function of the density $%
n(r,t)$ in the local and adiabatic approximation. Differentiating both sides
of this equation with respect to the charge density $n(r,t)$ one finds a
Dyson like equation for the interacting susceptibility $\chi _{int}=\frac{%
\delta n}{\delta V_{xc}}$: 
\begin{equation}
\chi _{int}(rt,r^{\prime }t^{\prime })=\frac{1}{\chi ^{-1}(rt,r^{\prime
}t^{\prime })-\frac{\delta (t-t^{\prime })}{|r-r^{\prime }|}%
-f_{xc}(rt,r^{\prime }t^{\prime })}\text{ }
\end{equation}
with a local exchange kernel $f_{xc}=\frac{dV_{xc}}{dn}$. The poles of $\chi
_{int}$ as a function of frequency provide information on the excited
states. The preceding equation requires as input the susceptibility $\chi $
of independent Kohn-Sham reference fermions 
\begin{equation}
\chi (\omega ,r,r^{\prime })=\sum_{i,j}\frac{n_{i}-n_{j}}{\omega
+i\varepsilon -(E_{i}-E_{j})}\phi _{i}^{\ast }(r)\phi _{i}(r^{\prime })\phi
_{j}(r)\phi _{j}^{\ast }(r^{\prime })\text{ }  \label{KSsusceptibility}
\end{equation}
where $\{\phi _{i}(r),E_{i}\}$, $i=1..N$ is a set of $N$ Kohn-Sham
eigenfunctions and energies. From eq (\ref{KSsusceptibility}) we see that $%
\chi $ is a sum of $\ \sim N^{2}$ terms that must be computed for a set of $%
N_{\omega }$ frequencies and with $r,r^{\prime }$ each ranging over $%
V/\Delta V$ cells in space, with $V$ the volume of the system and $\Delta V$
an elementary grid volume. Because the number of orbitals $N$ grows linearly
with the volume $V$, the calculation of $\ $Kohn-Sham susceptibility $\chi $
then requires $\sim N^{4}N_{\omega }$ operations, more than the remaining
operations needed to calculate the interacting susceptibility $\chi _{int}$,
the number of which grows with $N$ only as $\sim N^{3}N_{\omega }$.

{\it Accelerated Computation}. In LCAO schemes such as, for example, the
siesta approach \cite{siesta}, one develops the fermion operators in terms
of a set of orbitals $\{\psi _{a}(r)\}$ of finite range. This implies a
corresponding expansion for the Kohn-Sham density response function $\chi $
and one must now find 
\begin{equation}
\text{ }\chi _{abcd}(t)=\sum_{E,F}\psi _{a}^{\ast E}\psi _{b}^{E}\psi
_{c}^{\ast F}\psi _{d}^{F}\frac{n(E)-n(F)}{\omega +i\varepsilon -(E-F)}.
\end{equation}
Here the $\psi _{a}^{E}$ are eigenvectors of the Kohn-Sham hamiltonian in
the orbital basis and the indices of $\chi _{abcd}$ each range over $N$
orbitals (which we take to be real, so we drop the complex conjugation).
There appear to be far too many independent quantities $\chi _{abcd}$, but
for orbitals of finite range locality limits the distance between,
respectively, the orbitals ($a,c$) and ($b$,$d$) to a few atomic distances
which is also quite natural as these pairs of orbitals correspond to
coincident points in the continuum limit. Therefore, in a basis of localized
orbitals, only a subset of $\sim N^{2}$ susceptibilities $\chi _{abcd}$
contributes in the linear response equations, see \cite{BlaseOrdejon} for a
more detailed discussion.

The Kohn-Sham susceptibility of free fermions is just a product (or
convolution in frequency) of propagators, 
\begin{equation}
\chi _{abcd}(t)=-iG_{ab}(t)G_{cd}(-t)
\end{equation}
so its calculation is essentially finished once we know the propagator of
the free Kohn-Sham fermions. The determination of $G_{ab}(\omega )$ requires
of the order of $\sim N^{3}$ operations per frequency (even if the
eigenfunctions are already available) or $\sim NN_{\omega }$ operations for
a given set of indices $\{a,b,c,d\}$ at all frequencies, and these
operations totally dominate the low $\sim N_{\omega }\log N_{\omega }$
calculational cost of a fast convolution.

For this reason, and also because convolutions of Green functions require
great care to avoid finite size effects, we give here a way of calculating $%
\chi _{abcd}(\omega )$ that is comparable in overall speed, but more
accurate than a convolution. We decompose the susceptibility into positive
and negative frequency components, $\chi _{abcd}(\omega )=\chi
_{abcd}^{-}(\omega )+\chi _{cdab}^{-}(-\omega )$ with 
\begin{equation}
\chi _{abcd}^{-}(\omega )=\sum_{E<0<F}\frac{\rho _{ab}^{E}\rho _{cd}^{F}}{%
\omega -i\varepsilon -(E-F)}\text{ \ \ }
\end{equation}
where $\rho _{ab}^{E}=\psi _{a}^{E}\psi _{b}^{E}$ and rewrite $\chi
_{abcd}^{-}(\omega )$ as 
\begin{eqnarray}
\chi _{abcd}^{-}(\omega ) &=&-\sum_{E<0}\rho _{ab}^{E}G_{cd}^{+}(-\omega +E)%
\text{ \ }  \label{convolution} \\
G_{cd}^{+}(\omega ) &=&\sum_{F>0}\frac{\rho _{cd}^{F}}{\omega +i\varepsilon
-F}
\end{eqnarray}
\ where $G_{cd}^{\pm }(\omega )$ denotes the positive/negative frequency
part of the propagator. We also discretize the frequency axis to a finite
mesh of $N_{\omega }$ points which we choose to be uniform, for simplicity,
and we interpolate $G_{cd}^{+}(\omega )$ linearly between the mesh points.
Because $G_{cd}^{+}(\omega )$ varies on a frequency scale of $\varepsilon $
this interpolation introduces a relative error in $\chi _{abcd}^{-}$ of the
order of $\left( \Delta \omega /\varepsilon \right) ^{2}$ where $\Delta
\omega $ is the frequency spacing. The sum in eq (\ref{convolution}) can be
represented as a convolution, $\chi _{abcd}^{-}=\frac{1}{\pi }%
\mathop{\rm Im}%
G_{ab}^{-}\circledast G_{cd}^{+}(-\omega )$ that can be calculated much
faster, but the result is less accurate and only useful in conjunction with
an accelerated calculation of $G_{ab}^{\pm }$ itself.

After reconstituting $\chi _{abcd}$ from $\chi _{abcd}^{-}$ we have spent $%
\sim NN_{\omega }$ operations on $\chi _{abcd}$ for a given set of indices $%
\{a,b,c,d\}$. As mentioned before, only $\sim N^{2}$ combinations of such
indices enter into the equation of the interacting susceptibility, and so we
have indeed reduced the computational cost of the Kohn-Sham susceptibility
from $\sim N^{4}N_{\omega }$ to $\sim N^{3}N_{\omega }$ operations.

For a molecule bound to a slab of surface layers and similar systems that
involve thousands of orbitals, this speed up by a factor $N$ \ could turn
out to be essential for the feasibility of the computation.

{\it Test of the procedure.} To illustrate the quality of our approximate
procedure as a function of the number of orbitals $N$, the number of
frequency points $N_{\omega }$ and the smoothing parameter $\varepsilon $,
we choose a symmetric hamiltonian $h$ at random with a probability: 
\begin{equation}
prob(h)\sim \exp -N\text{ }\sum_{i,k=1..N}h_{ik}^{2}\text{ }
\end{equation}
from the so called ''orthogonal ensemble'' of random matrices. For $%
N\rightarrow \infty $, the density of eigenvalues of this matrix ensemble
tends to a semicircular distribution in the interval $(-1,1)$, see \cite
{Wigner}. We therefore choose units in which the band ranges between $(-1,1)$%
, set $E_{f}=0$ and calculate $\chi _{abcd}$ \ (i) exactly and (ii)
approximately in, respectively $\sim N^{2}N_{\omega }$ and $\sim NN_{\omega
} $ operations and compare our results.

With a reasonable choice of the values of $\varepsilon $ and the number of
mesh points $N_{\omega }$, the exact and approximate values of $\chi $
coincide to within the linewidth of the plot, as seen in figure (1). Figure
(2) confirms that the error scales like $\left( \Delta \omega /\varepsilon
\right) ^{2}$ with the mesh spacing. For $N=2000$, the approximate
calculation is about a thousand times faster than the exact one.

{\it Conclusion}. In summary, we have described an accelerated procedure for
computing the charge response of Kohn-Sham reference fermions in order to
facilitate the calculation of TDDFT linear response for systems that are
intrinsically of large size, such as molecules bound to solid surfaces. Our
speedup from $\ N^{4}N_{\omega }$ $\ $to $N^{3}N_{\omega }$ operations for
calculating the subset of elements of $\chi _{abcd}$ that contribute in
linear response comes at the price of a finite precision of the order of $%
\sim \left( \frac{\Delta \omega }{\varepsilon }\right) ^{2}$ where $\Delta
\omega $ is the frequency spacing.

A reduction from $N^{4}N_{\omega }$ to $N^{3}N_{\omega }$ operations in
TDDFT linear response was achieved before by limiting the density response
to a finite manifold of fit functions \cite{CasidaGisbergen}. Because our
procedure is very simple it might be combined with such or similar methods
to further reduce the growth of operations with $N$.

We expect our algorithm to facilitate the study, by TDDFT linear response,
of optical properties of hybrid systems such as molecules docked on
surfaces, systems that are difficult to treat by existing computational
methods and which exhibit many interesting features that deserve further
study.

{\bf Acknowledgements}. I wish to thank Andrei Postnikov (Osnabr\"{u}ck) for
useful discussions and continued correspondence and Xavier Blase (Lyon) for
sending his work on TDDFT linear response prior to publication. I am
indebted to Daniel Sanchez-Portal (San Sebastian) for discussions on the
siesta code and to Mark Casida (Grenoble) for interesting comments on his
approach\ to linear response. Computer time was provided by IDRIS (Orsay)
and Regatta (Bordeaux) and financial support by GDR-DFT. I thank Alois
W\"{u}rger for discussions on the manuscript.

{\bf Figure Captions:}

Figure (1) : Re$\chi _{abcd}$, Im$\chi _{abcd}$ as a function of frequency
for $N=2000$ orbitals, $\varepsilon =0.03$, $N_{\omega }=256$ frequencies
and for a random choice of $\{a$,$b$,$c$,$d\}$. The difference between the
exact and the approximate values of $\chi _{abcd}$ is smaller than the
linewidth of the figure, therefore one can only see two distinct curves.

Figure (2) The relative error in $\chi _{abcd}$ for $N=2000$ orbitals and $%
\varepsilon =0.03$ as a function of the number of frequency points $%
N_{\omega }$. The relative error is defined as the ratio of the root mean
squares of error and signal and, as expected, it scales as $\sim N_{\omega
}^{-2}$ or $\sim \left( \Delta \omega \right) ^{2}$.

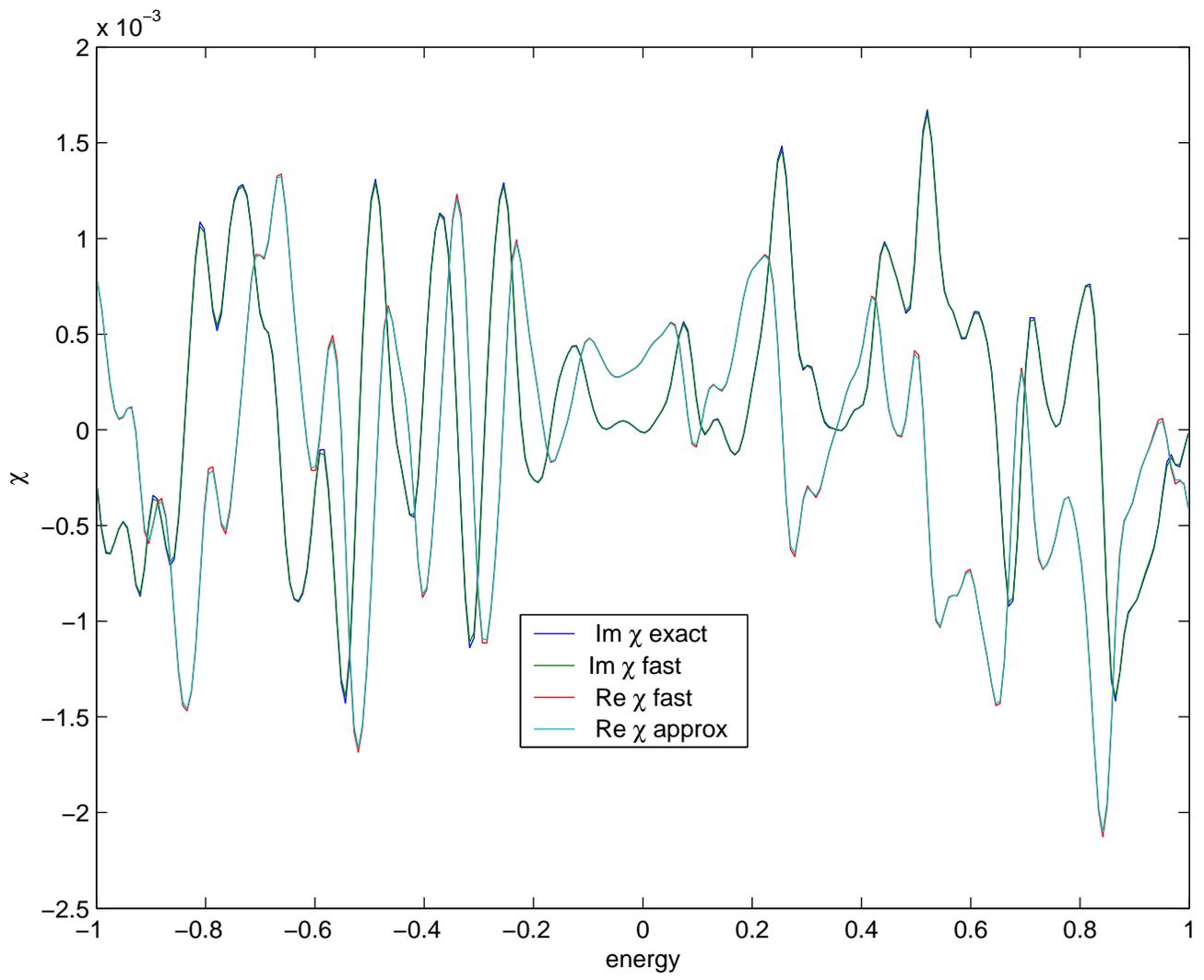
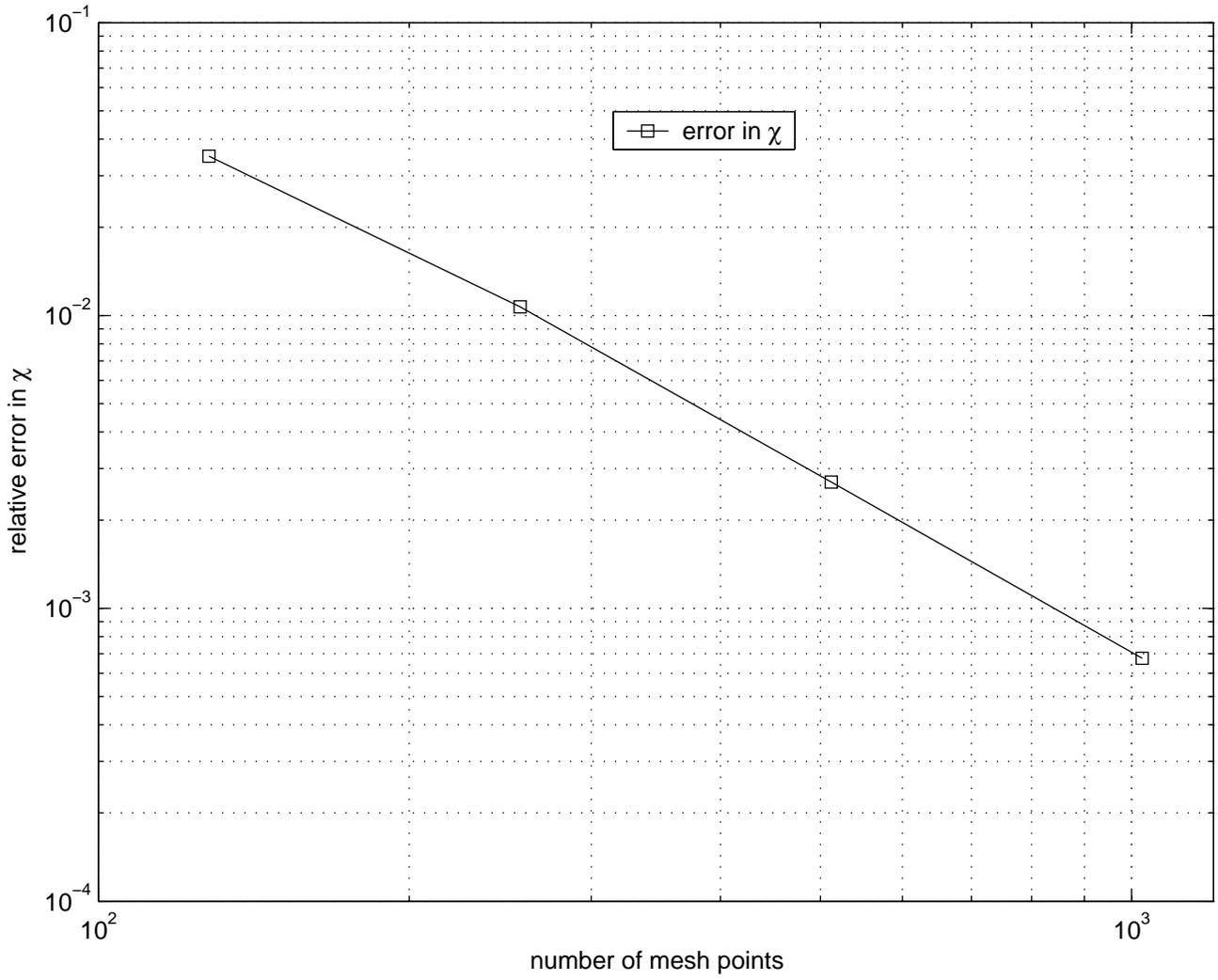
\end{document}